\documentclass[final,twocolumn,12pt]{elsarticle}

\usepackage{amssymb}
\usepackage{comment}
\usepackage{upgreek}
\usepackage{amsmath}
\usepackage{esint}
\usepackage{url}
\usepackage{nicematrix}
\usepackage[hidelinks]{hyperref}
\usepackage{lineno}
\usepackage{graphicx}
\usepackage{bm}
\usepackage{multirow}

\newcommand{\rv}[1]{\textcolor{black}{#1}}

\journal{Sensors and Actuators A. Physical}

\begin{document}


\begin{frontmatter}

\title{Optimized waveguides for mid-infrared lab-on-chip systems: A rigorous design approach}


\author[telma]{Antonia Torres-Cubillo\corref{atc}}
\ead{atc@ic.uma.es}
\cortext[atc]{Corresponding author}
\author[uulm,art]{Andrea Teuber}
\author[telma,ibima]{Robert Halir}
\author[uulm,hahn]{Boris Mizaikoff}


\affiliation[telma]{organization={Telecommunication Research Institute (TELMA), University of Malaga},
            city={29010 Malaga},
            country={Spain}
}

\affiliation[uulm]{organization={Institute of Analytical and Bioanalytical Chemistry, University of Ulm},
            city={Ulm 89081},
            country={Germany}
}

\affiliation[art]{organization={Art Photonics},
            city={Ulm 89077},
            country={Germany}
}

\affiliation[ibima]{organization={IBIMA-BIONAND},
            addressline={Malaga Tech Park}, 
            city={29010 Malaga},
            country={Spain}
}

\affiliation[hahn]{organization={Hahn-Schickard},
            city={Ulm 89077},
            country={Germany}
}


\begin{abstract}

Mid-infrared absorption spectroscopy is a well-established technique for non-destructive quantitative molecular analysis. Waveguide-integrated sensors provide a particularly compact solution operating with reduced sample volumes while exhibiting exquisite molecular selectivity, sensitivity, and ultra-low limits of detection. Recent advances in mid-infrared technologies along with the integration of on-chip sources, detectors and microfluidics, have brought mid-infrared lab-on-chip systems closer to reality. A variety of material platforms has been proposed for the implementation of such systems. However, the lack of a consistent waveguide design approach renders a fair comparison between different alternatives -- and a deliberate material selection -- challenging, limiting the development of optimized on-chip spectroscopic devices. In the present study, a systematic waveguide design approach has been developed, facilitating evanescent field absorption-based sensing, in particular for aqueous analytes. Our strategy enables a rigorous comparison of several state-of-the-art thin-film waveguides using parametric expressions to predict the achievable limits of detection of the sensing system, while indicating optimum waveguide dimensions and absorption pathlengths, pivotal for the development of next-generation mid-infrared lab-on-chip devices.  

\end{abstract}


\end{frontmatter}


\section{Introduction} \label{sec:intro}

\begin{figure}[tb]
\centering
\includegraphics[width=1\linewidth]{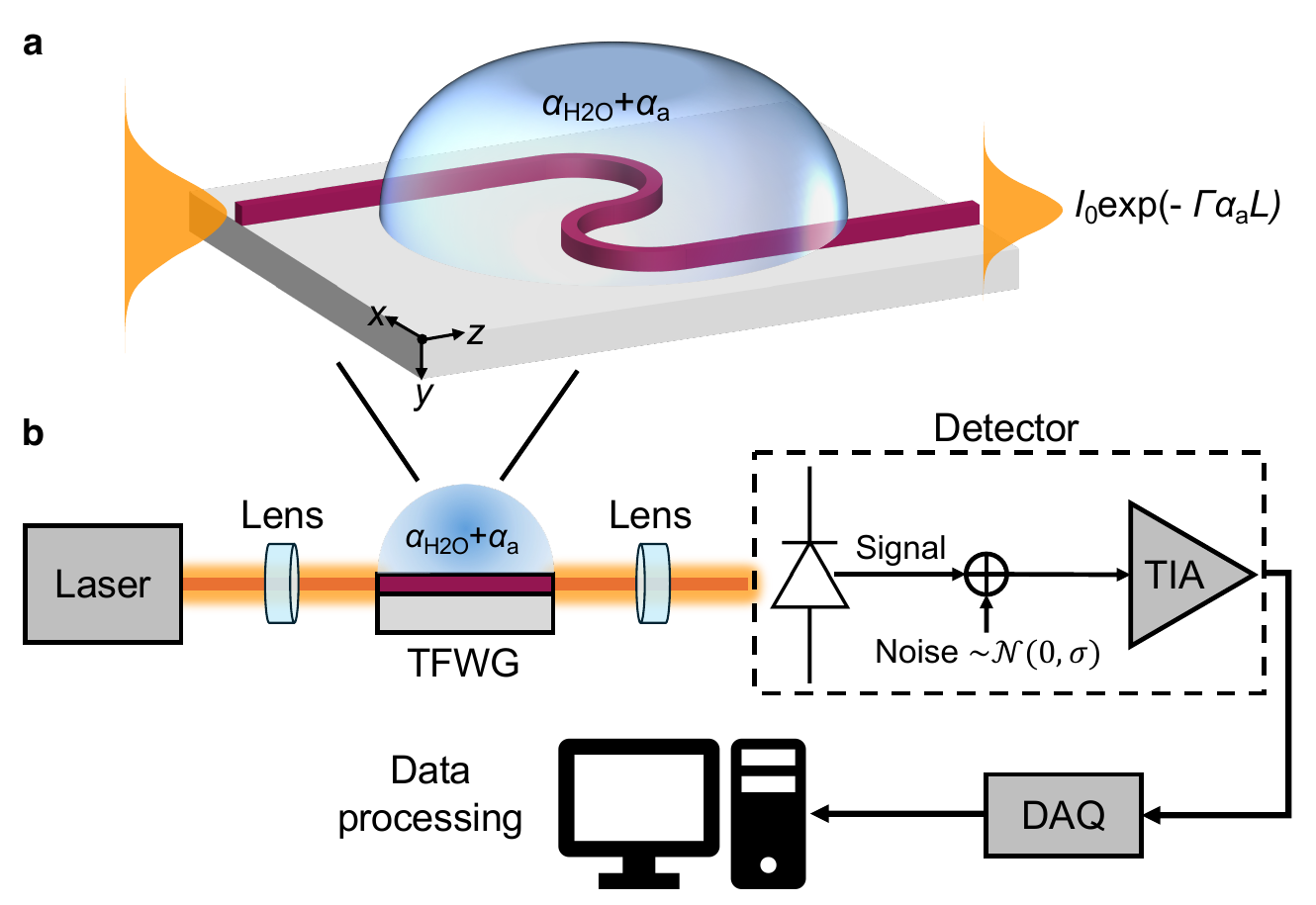}
\caption{(a) Photonic integrated waveguide sensor for MIR absorption spectroscopy. (b) Schematic of the standard MIR spectroscopic setup considered for the determination of the LOD. White noise is added at the input of the transimpedance amplifier.}
\label{fig:wg_setup}
\end{figure}

Mid-infrared (MIR) absorption spectroscopy is a powerful and reliable analytical tool, which provides inherent molecular specificity and enables the quantitative and qualitative detection of a wide variety of analytes \cite{Bec2020,Selvaraj2020,Fu2022}. Waveguide-based MIR absorption sensors, as schematized in Fig. \ref{fig:wg_setup}(a), confine and manipulate light in photonic waveguides, and rely on the evanescent field sensing principle. The absorption of the guided mode depends on the interaction of its evanescent field with the sample. Quantitatively, a sample may be described as a target analyte diluted into a solvent, where light propagation is governed by the Beer-Lambert law:
\begin{equation}
    I=I_0\exp(-\Gamma\alpha_\mathrm{a}L),
\end{equation}
where $I$ is the recorded output intensity, $I_0$ is the background intensity, $\Gamma$ is the light confinement factor in the sample \cite{Robinson2008}, $\alpha_\mathrm{a}$ is the absorption coefficient of the target analyte and $L$ is the interaction length. The background intensity $I_0=I_\mathrm{in}\exp(-\alpha_\mathrm{wg}L)$ depends on the input intensity $I_\mathrm{in}$, and waveguide loss $\alpha_\mathrm{wg}$, which accounts for all absorption phenomena not caused by the analyte, such as absorption by the solvent, leakage to a higher-index substrate or losses in the waveguide materials. These sensors can be extremely compact and work with reduced sample volumes, while still exhibiting an exquisite performance \cite{Ma2020}. Moreover, on-chip integration favors large-scale production and facilitates compatibility with complementary metal oxide semiconductor (CMOS) fabrication techniques \cite{Royo2013}, thus reducing the costs of components for intelligent sensor networks or disposable point-of-care devices \cite{Kazanskiy2020}. Indeed, recent advances in the monolithic integration of MIR sources \cite{Wang2022,Karnik2024} and detectors \cite{Dai2022,Giparakis2024}, along with microfluidics \cite{Testa2022}, are bringing lab-on-chip (LOC) devices closer to reality \cite{Hinkov2022}. This could transform areas like early medical diagnosis by enabling in-situ detection of critical biomarkers in real-time, ideally performed by untrained personal, while delivering precise results. 

For trace-gas detection in the MIR, suspended \cite{Ottonello2020,Sanchez2021}, slot \cite{Yallew2023,Zhang2023} and hollow \cite{Wu2020,Hlavatsch2023} waveguides have been proposed, in many cases reporting confinement factors beyond free-space \cite{Vlk2021,Pi2023}. Alternatively, analyzing aqueous analyte solutions is of substantial interest, as water – besides being an exceptionally strong IR-absorber – is the main component of bodily fluids such as saliva, urine, or tears. However, these solutions pose specific challenges. Compared to gases, they require waveguides with superior mechanical robustness, and, therefore, solid thin-film waveguides (TFWG) are preferred. These waveguides usually exhibit modest sensitivities, as only a small fraction of the evanescent field is in contact with the sample, i.e. $\Gamma$ is small. Again, the background itself is strongly absorbing and exhibits spectral overlaps with some organic components \cite{Fomina2023}. Despite these drawbacks, very low limits of detection (LOD) have been reported via TFWG implementations fabricated from different material platforms \cite{Mittal2020}. Gallium arsenide (GaAs) slab waveguides achieved a LOD of 56.9 mM for ammonium perchlorate \cite{Sieger2016}. A surface-enhanced GaAs platform was used for direct broadband spectroscopy, achieving a LOD of 3500 ppb of AFB1 mycotoxin \cite{Haas2020}. Chalcogenide waveguides combined with paper microfluidics have been demonstrated for absorption spectroscopy of isopropyl alcohol in water as low as 20\% \cite{Mittal2018}. Ge-Sb-Se (Se6/Se2) waveguides functionalized with a hydrophobic polymer could reach an estimated LOD of 26 ppb of toluene in water \cite{Baudet2017}. The protein aggregates of a 900 µM bovine serum albumine (BSA) solution could be precisely distinguished from the spectra provided by a germanium on silicon (GOS) multimode waveguide \cite{Mittal2020Ge}. A rib GOS waveguide was topped with a mesoporous silica cladding and achieved a LOD of 7 ppm of toluene in water \cite{Beneitez2020}. A 5 µm thick polycristalline diamond (PCD) waveguide on aluminium nitride obtained a LOD of 0.05\% for acetone \cite{Forsberg2021}. By enhancing the surface of a diamond waguide with graphene, taurine concentrations below 0.4w\% were detected \cite{Teuber2023}. 

As the origin of the analytical signal, a well-designed sensing waveguide is the essential component of a high-performance integrated MIR absorption spectroscopy system, ensuring intimate and reproducible interaction of photons and sample molecules. Despite the wide range of proposed platforms and geometries, a consistent design approach for MIR TFWGs has not yet been established. Consequently, most sensors exhibit different characteristics and have been evaluated using different analytes, thus limiting a direct comparison. In turn, this renders devising a clear route towards fully optimized LOC systems challenging. Here, we propose a systematic design approach for MIR waveguides for aqueous analytes based on the insight provided by an analytical expression for the limit of detection of absorption-based spectroscopy systems. Waveguide sensitivity and losses are calculated via rigorous electromagnetic simulations and combined into a meaningful figure-of-merit. The achievable LOD for each waveguide is evaluated yielding a quantitative parameter for overall performance. This enables, for the first time, to establish a fair comparison between the sensing capabilities of four representative TFWG platforms, thereby facilitating platform selection and waveguide design for optimized sensing devices. 

\section{Results} \label{sec:results}

\subsection{System-level model} \label{subsec:sytem}

For our analysis, the spectroscopic setup shown in Fig. \ref{fig:wg_setup}(b) is considered to determine the limit of detection, i.e. the smallest amount of analyte-induced absorption that can be reliably detected. The LOD is usually defined as $\mathrm{LOD}=3\sigma/S$, where $\sigma$ is the noise present in the output signal and $S$ is the sensitivity of the output signal to changes in the concentration of the analyte \cite{Loock2012}. Light from a laser is focused via a lens on the waveguide with length $L$, travels through it and interacts with the analyte. Afterwards, the light beam is collected with a second lens and illuminates an appropriate photodetector, resulting in a photovoltage or photocurrent, which is the output signal. To account for the different noise sources, including shot noise, thermal noise from the transimpedance amplifier (TIA) and quantization noise from the DAQ, we added an equivalent white Gaussian noise to the photocurrent with a standard deviation:
\begin{equation}
\sigma=\sqrt{\sigma_\mathrm{shot}^2+\sigma_\mathrm{TIA}^2+\sigma_\mathrm{DAQ}^2}
\label{eq:sigma}
\end{equation}
where $\sigma_\mathrm{shot}$, $\sigma_\mathrm{TIA}$ and $\sigma_\mathrm{DAQ}$ are the standard deviations of shot, TIA and quantization noise respectively \cite{Molina2019}. Detailed parametric expressions for each noise contribution can be found in Supplementary Material (Supp. Mat.) S1. The LOD is then given by: 
\begin{equation}
    \mathrm{LOD}=\frac{3\sigma}{S_0\Gamma L e^{-\alpha_\mathrm{wg}L}}
    \label{eq:LOD}
\end{equation}
Here, $L$ is the sensing pathlength, $\alpha_\mathrm{wg}$ is the waveguide loss defined in Section \ref{sec:intro} (Introduction), and $S_0$ is a constant factor that accounts for system-level parameters (see Supp. Mat. S1). This expression can be particularized for any specific system and readily extended to incorporate further noise sources. From Eq. \eqref{eq:LOD} it is evident that both the confinement factor $\Gamma$ and the waveguide loss $\alpha_\mathrm{wg}$ are critical for determining the LOD. It is also observed that an optimum sensor length exists as a result of a trade-off between accumulating enough signal change due to interaction with the analyte ($\Gamma L$) and avoiding excessive signal attenuation due to losses ($e^{-\alpha_\mathrm{wg}L}$). \rv{In fact, such optimum length can be calculated as $L_\mathrm{opt}=1/\alpha_\mathrm{wg}$ \cite{Molina2019}.}

\subsection{Waveguide platforms} \label{subsec:platforms}

\begin{figure*}[tb]
\centering
\includegraphics[width=0.9\linewidth]{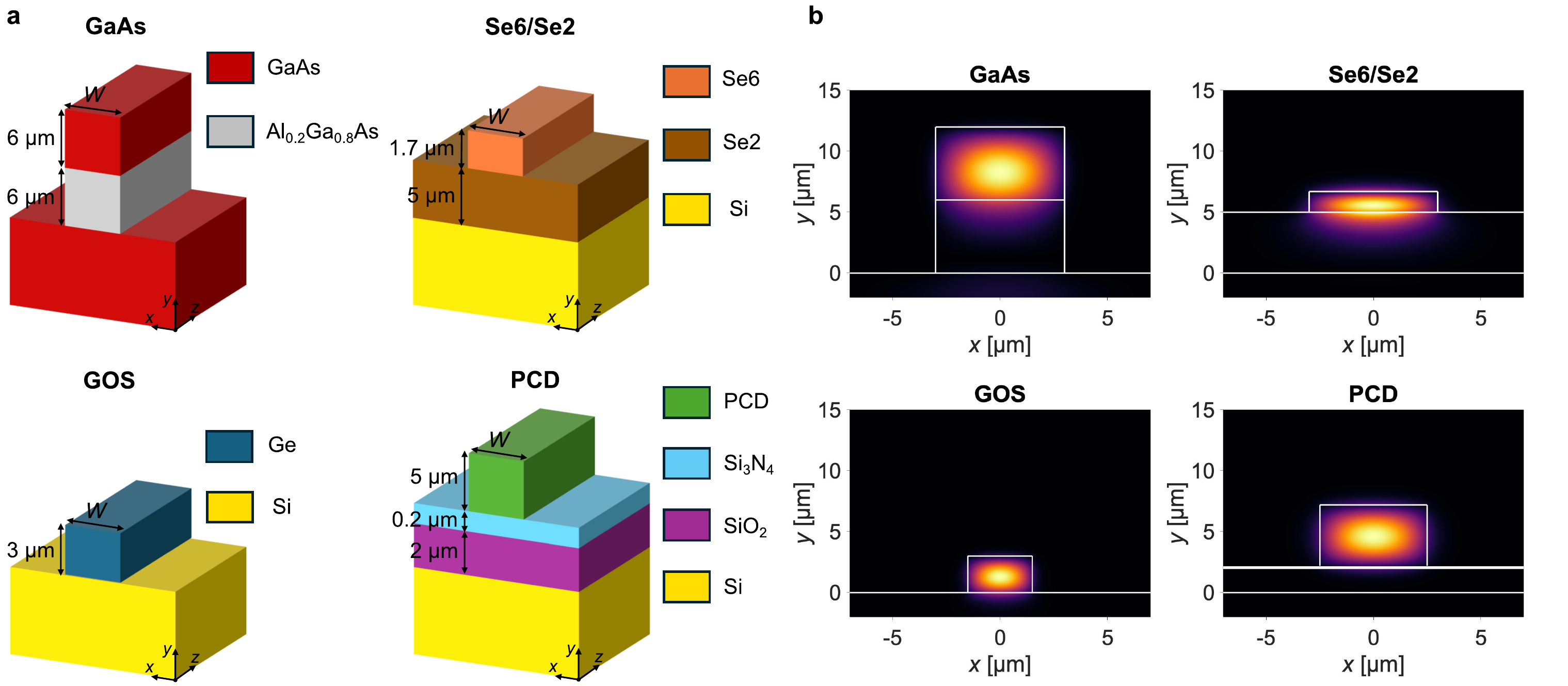}
\caption{Overview of studied platforms. (a) Schematic representation of the four studied waveguide platforms: gallium arsenide (GaAs), chalcogenide glass (Se6/Se2), germanium on silicon (GOS) and polycrystalline diamond (PCD). Layer thicknesses are not to scale. (b) Y-component of the electric field distribution ($|E_\mathrm{y}(x,y)|$) of the fundamental TM mode supported by each of the studied waveguides.}
\label{fig:platforms}
\end{figure*}

\begin{table}
\begin{center}
\caption{Values for the real (n) and imaginary (k) parts of the complex refractive index of the materials considered in this work at $\lambda=6\,\mathrm{\upmu m}$. }
\label{tab:nk}
\begin{NiceTabular}{ c c c c}
\hline
Material & $n$ & $k$ & Ref.\\
\hline
GaAs & 3.3& 0.0004695 & \cite{Rakic1996,Skauli2003} \\
$\mathrm{Al_{0.2}Ga_{0.8}As}$ &3.2000	& 0.0024 & \cite{Rakic1996} \\
Se6	& 2.8 & 	0	& \cite{Nvemec2014} \\
Se2 & 2.4 &	0	& \cite{Nvemec2014} \\
Si	& 3.4698 &	0.0000017	& \cite{Shkondin2017} \\
Ge	& 3.9671 &	0 &	\cite{Amotchkina2020} \\
PCD	& 2.4214	& 0.000553	& \cite{Dore1998} \\
$\mathrm{Si_3N_4}$ &	2.2236	& 0.032326	& \cite{Kischkat2012} \\ 
$\mathrm{SiO_2}$	& 1.2713	& 0.0016067	& \cite{Kischkat2012} \\
$\mathrm{H_2O}$	& 1.265 &	0.107	& \cite{Hale1973} \\
\hline
\end{NiceTabular}
\end{center}
\end{table}

Four different TFWGs are considered in the following, which are commonly employed for MIR spectroscopic analysis of liquid analytes \cite{Mittal2020}, namely GaAs \cite{Sieger2016}, Se6/Se2 \cite{Baudet2017}, GOS \cite{Mittal2020Ge} and PCD \cite{Haas2018}. An overview of the selected platforms is presented in Fig. \ref{fig:platforms}(a) where the thickness of the layers of the different materials is specified. Two representative central wavelengths ($\lambda$), 6 µm (1667 cm$^{-1}$, amide vibrational band) and 10 µm (1000 cm$^{-1}$, carbohydrate vibrational band) are considered \rv{as significant examples} during this fundamental study. \rv{However, the following workflow is independent on the specific wavelength, and results could be replicated at any other band by properly modifying the models of the TFWG materials.} The analysis is restricted to single-mode (SM) waveguide dimensions, to ensure a controlled photon-analyte interaction, and focuses on the transversal magnetic (TM) polarization to enhance compatibility with quantum cascade laser (QCL) sources. Given the focus on aqueous analytes, pure water is considered as the waveguide cladding. The values for the complex refractive indexes ($n+\mathrm{i}k$) of all waveguide materials are compiled in Table \ref{tab:nk} for the 6 µm working wavelength. It should be noted that water exhibits the highest absorption coefficient among the materials within these waveguide configurations. In the following, the critical waveguide parameters ($\Gamma,\alpha_\mathrm{wg}$) for these platforms are rigorously calculated. 

\subsection{Confinement factor and waveguide losses} \label{subsec_fom} 

\begin{figure*}[tb]
\centering
\includegraphics[width=0.8\linewidth]{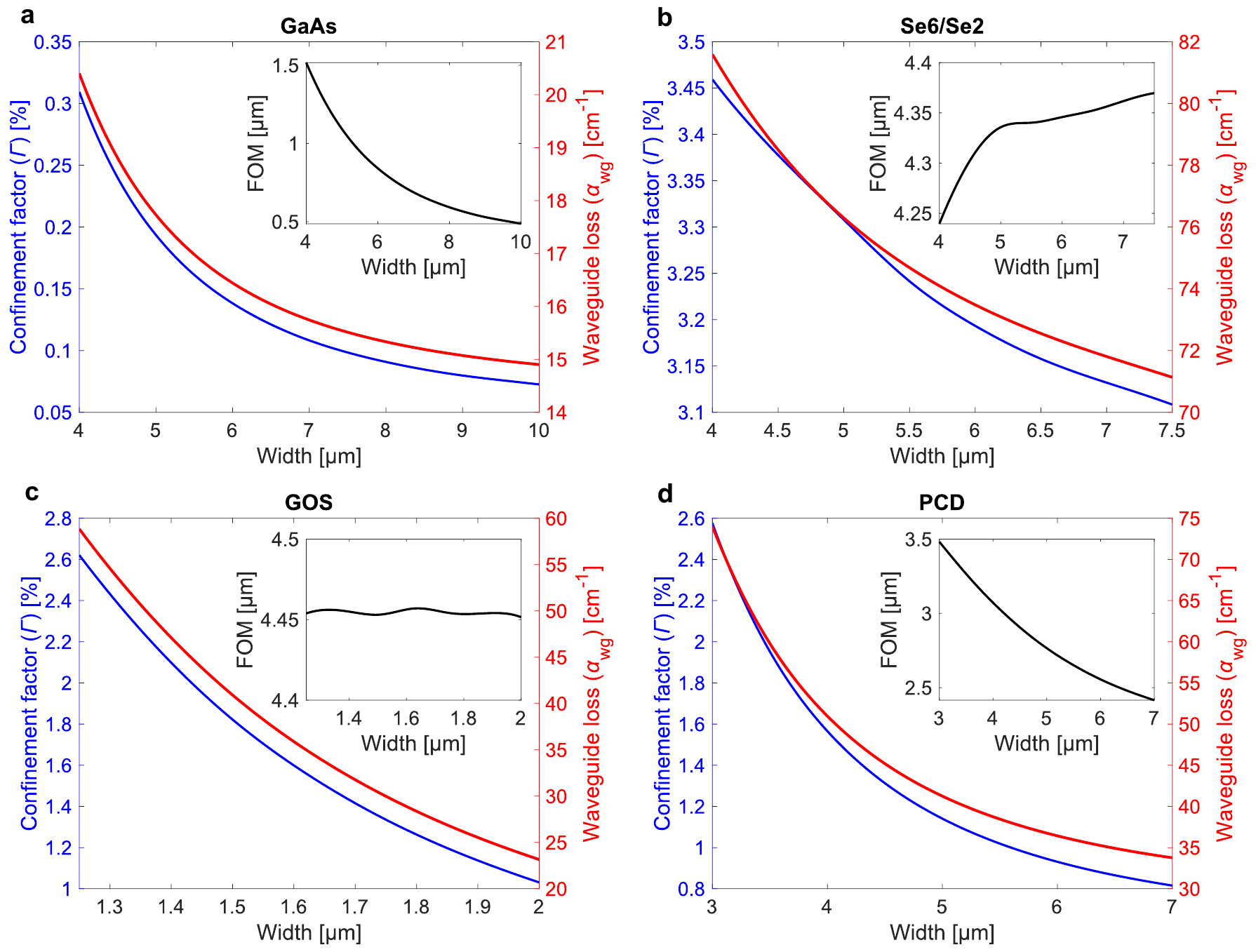}
\caption{Performance of the single-mode TFWGs. Confinement factor ($\Gamma$) and waveguide loss ($\alpha_\mathrm{wg}$) for (a) GaAs/AlGaAs, (b) Se6/Se2, (c) GOS and (d) PCD waveguides for widths within the single-mode regime. The figure of merit (FOM) of each waveguide is shown in the insets.}
\label{fig:FOM}
\end{figure*}

\begin{table}
\begin{center}
\caption{Penetration depths of the fundamental TM mode for the different platforms.}
\label{tab:dp}
\begin{NiceTabular}{c| c c c c}
\hline
Platform &	GaAs &	Se6/Se2	& GOS &	PCD\\
$d_\mathrm{p}$\,[µm] &	0.4 & 	0.45 & 0.3 &	0.5 \\
\hline
\end{NiceTabular}
\end{center}
\end{table}

\rv{By performing Finite Element Method (FEM) simulations \cite{Jin2015,FemSIM}}, the SM regime of each waveguide is calculated as the width span in which just the fundamental TM mode, whose electric field distribution ($|E_\mathrm{y}(x,y)|$) is shown in Fig. \ref{fig:platforms}(b), is supported. Afterwards, the confinement factor ($\Gamma$) is evaluated across these widths, yielding the results shown in Fig. \ref{fig:FOM}. The corresponding simulation techniques are \rv{further} described in Section \ref{sec:methods} (Methods). For comparison, the penetration depth ($d_\mathrm{p}$) of the TM mode for each platform is also calculated, as this metric is often applied to compare TFWG platforms. The obtained values are given in Table \ref{tab:dp}. Furthermore, the total waveguide loss coefficient ($\alpha_\mathrm{wg}$) is considered as:
\begin{equation}
    \alpha_\mathrm{wg} = \Gamma\alpha_\mathrm{H2O}+\alpha_\mathrm{int}
    \label{eq:awg}
\end{equation}
where $\alpha_\mathrm{H2O}$ are water absorption losses and $\alpha_\mathrm{int}$ are the intrinsic losses of the waveguide, which account for absorption by the platforms materials and leakage to a higher-index substrate. The contribution of scattering losses due to surface roughness, which are dependent on the defects and imperfections introduced in the waveguide geometry during the fabrication process, is neglected here. According to Payne-Lacey model \cite{Lacey1990}, the scattering loss coefficient is inversely proportional to the cubed wavelength, hence, this can be disregarded in the MIR regime. The resulting propagation losses, calculated as described in Section \ref{sec:methods} (Methods), are shown in Fig. \ref{fig:FOM}. To facilitate the comparison between the different alternatives, \rv{the trade-off between confinement factor} and losses \rv{is evaluated by} a single performance-based figure-of-merit (FOM) defined as:
\begin{equation}
    \mathrm{FOM}=\frac{\Gamma}{\Gamma\alpha_\mathrm{H2O}+\alpha_\mathrm{int}},
    \label{eq:FOM}
\end{equation}
which is shown in the insets of Fig. \ref{fig:FOM}(a)-(d).

\subsection{Limit of detection} \label{subsec:lod}

\begin{table*}
\begin{center}
\caption{Selected waveguide widths for the different platforms and associated performance metrics at $\lambda=6$ µm.}
\label{tab:width}
\begin{NiceTabular}{c c c c c c}
\hline
Platform & $W$\,[µm] & $\Gamma$\,[\%] & $\Gamma\alpha_\mathrm{H2O}$\,[cm$^{-1}$] & $\alpha_\mathrm{int}$\,[cm$^{-1}$]& FOM\,[µm]\\
\hline
GaAs & 7 & 0.11 & 2.43 & 13.32 & 0.69 \\
Se6/Se2 & 5.75 & 3.22 & 72.06 & 1.99 & 4.34 \\
GOS & 1.62 & 1.55 & 34.75 & 0.042 & 4.46 \\
PCD & 5 & 1.14 & 25.57 & 15.69 & 2.77 \\
\hline
\end{NiceTabular}
\end{center}
\end{table*}

\begin{figure}[tb]
\centering
\includegraphics[width=0.99\linewidth]{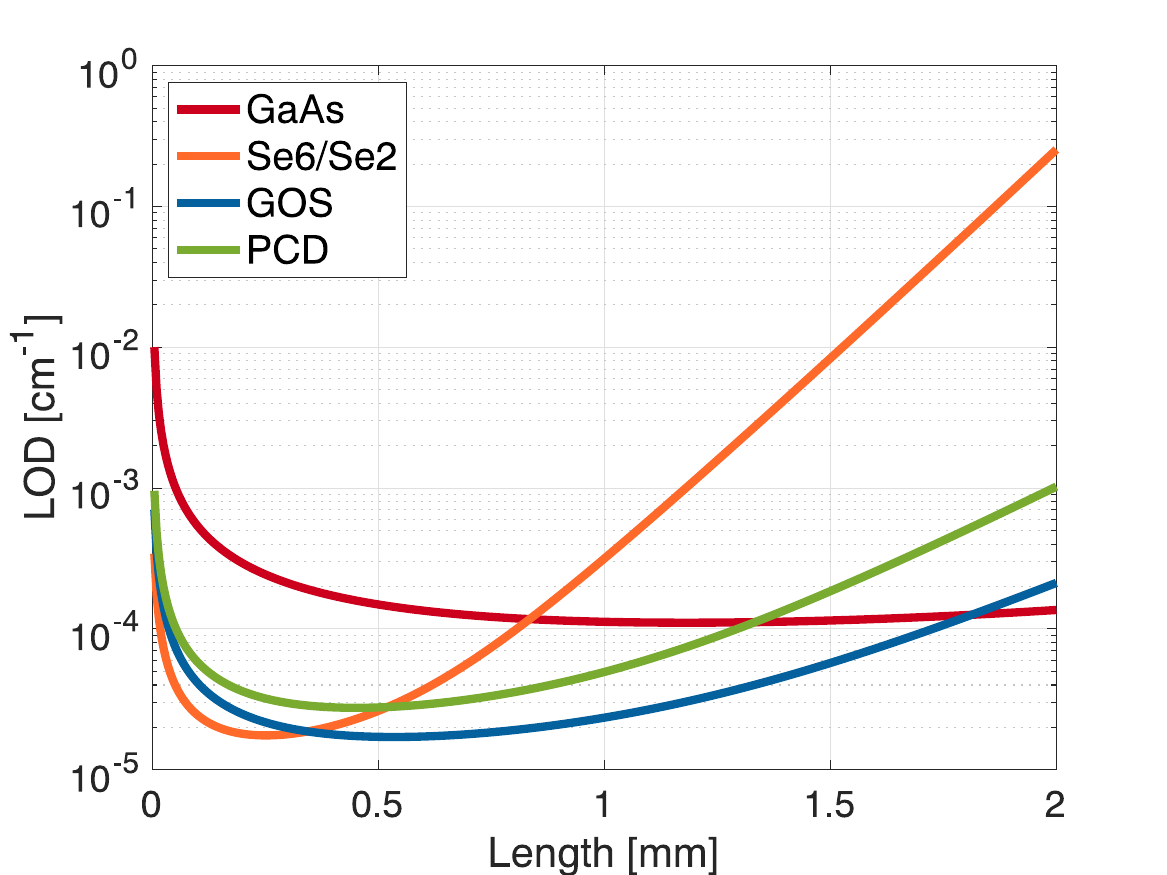}
\caption{Evaluated LOD as a function of the sensing pathlength for the different platforms. For each waveguide there is an optimum working region.}
\label{fig:LOD}
\end{figure}

\begin{table}
\begin{center}
\caption{Estimated optimum sensing pathlength and achieved LOD for the different evaluated platforms at $\lambda=6$ µm.}
\label{tab:LOD6}
\begin{NiceTabular}{c c c}
\hline
Platform & $L_\mathrm{opt}$\,[mm] & LOD\, [10$^{-5}$cm$^{-1}$]\\
\hline
GaAs & 1.8 & 11.1 \\
Se6/Se2	& 0.25 & 1.75 \\
GOS & 0.54 & 1.51 \\
PCD & 0.45 & 2.75 \\
\hline
\end{NiceTabular}
\end{center}
\end{table}

To study the limit of detection, a fixed width ($W$) for each waveguide was selected. Considering the trade-off between sensitivity and losses (see Fig. \ref{fig:FOM}(a)-(d)), a width in the center of the SM regime was chosen as a compromise solution, resulting in the widths and associated performance metrics summarized in Table \ref{tab:width}. The LOD of the different waveguides as a function of their sensing length is evaluated  via Eq. \eqref{eq:LOD}, where $\sigma$ and $S_0$ are given by a set of parameters for an exemplary -- and typical -- MIR setup: a MIRcat 2100 QCL source (DRS Daylight Solutions, USA), a PVI-4TE-10.6/MIP-10-1M-F-M4 amplified detector (Vigo System S.A., Poland) and a NI USB-6002 data acquisition board (National Instruments Corp., USA). The resulting LODs are presented in Fig. \ref{fig:LOD}. \rv{For the exact assumed experimental conditions and a discussion on the relative impact of the different noise sources on the overall limit of detection, see Supp. Mat. S1}. It should be noted that the LOD is expressed in absorption coefficient units (cm$^{-1}$), providing a general metric which is independent on the target analyte. Given an analyte of interest, its concentration can be calculated from the absorption coefficient as $C=\log(e)\alpha_\mathrm{a}/\varepsilon$, where $\varepsilon$ is the decadic molar absorption coefficient. For each TFWG platform, there is an optimum length at $\alpha_\mathrm{wg} L_\mathrm{opt}\approx1$, which is in the range of 1 mm and below. Here, it is worth noticing that shape and relative position of the curves depends only on the waveguide parameters $\Gamma$ and $\alpha_\mathrm{wg}$, and a change in system components and configuration would only result in scaling by the factor $\sigma/S_0$ (see Eq. \eqref{eq:LOD}). \rv{In fact, the optimum achievable LOD can be expressed in terms of the FOM of the waveguide, as}
\rv{
\begin{equation}
    \mathrm{LOD}(L_\mathrm{opt})=\frac{3\sigma}{(S_0/e)\mathrm{FOM}}.
    \label{eq:LODopt}
\end{equation}}
The estimated optimum LODs are given in Table \ref{tab:LOD6}. The studies discussed so far were replicated for the 10 µm wavelength regime. To ensure adequate waveguiding conditions in all platforms, the thicknesses of the guiding and buffer layers were scaled \rv{with wavelength, i.e., by a factor 10/6, so that the electrical size remains constant \cite{Saleh2019}.} Table \ref{tab:LOD10} shows the optimum LOD estimation for the wavelength of 10 µm. PCD waveguides have been excluded because they are unsuitable for this regime due to the excessive increase in the absorption of $\mathrm{Si_3N_4}$ and $\mathrm{SiO_2}$. Detailed information and intermediate results can be found in Supp. Mat. S2. 

\begin{table}
\begin{center}
\caption{Estimated optimum sensing pathlength and achieved LOD for the different evaluated platforms at $\lambda=10$ µm.}
\label{tab:LOD10}
\begin{NiceTabular}{c c c}
\hline
Platform & $L_\mathrm{opt}$\,[mm] & LOD\, [10$^{-5}$cm$^{-1}$]\\
\hline
GaAs & 3.26 & 4.86 \\
Se6/Se2	& 0.84 & 0.53 \\
GOS & 3.41 & 0.49 \\
\hline
\end{NiceTabular}
\end{center}
\end{table}

\subsection{Curvature losses} \label{subsec:curves}

\begin{figure*}[tb]
\centering
\includegraphics[width=0.7\linewidth]{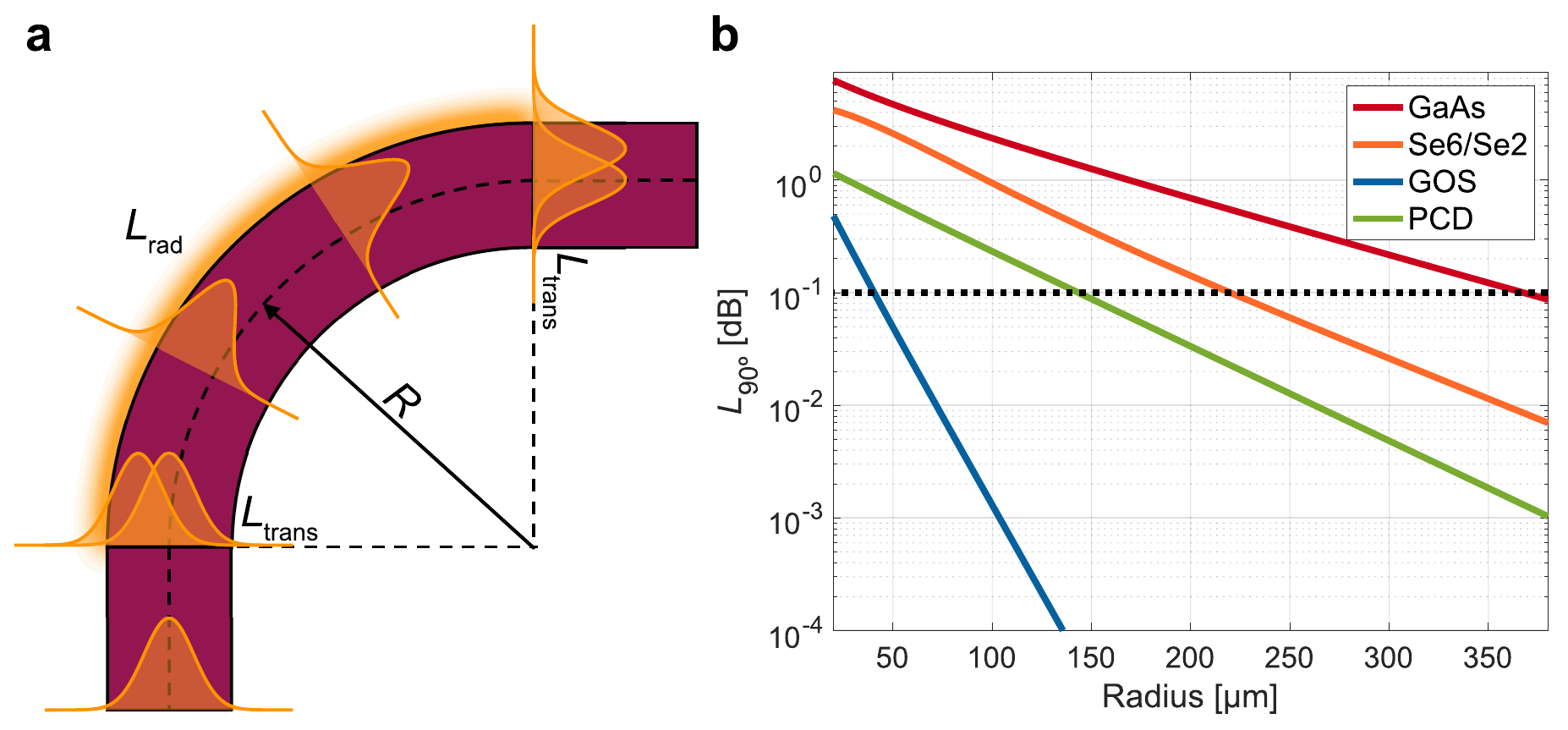}
\caption{(a) Generic 90$^\circ$ bent waveguide. Losses come from radiation in the curve and the mode mismatch in the transition between straight and bent sections. (b) Losses induced by a  90$^\circ$ bend for the different platforms at $\lambda=6$ µm.}
\label{fig:curves}
\end{figure*}

\begin{table}
\begin{center}
\caption{Bent radii which yield total bend losses below 0.1 dB for the different platforms and wavelengths. The contributions of radiation and transition losses are specified.}
\label{tab:L90}
\begin{NiceTabular}{c c c c c}
\hline
Platform & $\lambda$\,[µm] & $R$\,[µm] & $L_\mathrm{rad}$\,[dB] & $L_\mathrm{trans}$\,[dB] \\
\hline
\multirow{2}{*}{GaAs} & 6	& 369.1	& 0	& 0.05 \\
& 10 & 686.3 &	0 &	0.05 \\
\multirow{2}{*}{Se6/Se2} & 6 &	221.2	& 0.003 &	0.05\\
& 10	& 227.3	& 0.05	& 0.02 \\
\multirow{2}{*}{GOS} & 6 & 40.5	& 0.004 & 0.04 \\
& 10	& 58.7	& 0.003	& 0.04 \\
PCD & 6 &	144.4	& 0	& 0.05 \\
\hline
\end{NiceTabular}
\end{center}
\end{table}

Even though the obtained optimum sensing pathlengths can be implemented with straight waveguides, bends may be necessary for on-chip routing and the implementation of more complex sensing architectures. The loss introduced by a 90$^\circ$ bend (as shown in Fig. \ref{fig:curves}(a)) is defined as:
\begin{equation}
    L_{90}=L_\mathrm{rad}+2L_\mathrm{trans},
\end{equation}
where $L_\mathrm{rad}$ is the radiation loss along the 90$^\circ$ curve and $L_\mathrm{trans}$ is the mode mismatch loss in the transition between a straight and a bent waveguide. The evolution of $L_{90}$ with the bend radius at $\lambda=6$ µm for the different platforms is shown in Fig. \ref{fig:curves}(b). The same representation for the 10 µm wavelength is given in Fig. S5 of Supp. Mat. S2. Table 6 provides the bend radius for which $L_{90}<0.1$ dB and the individual contributions of radiation and transition losses.

\section{Discussion} \label{sec:disc}

\subsection{Waveguide figure-of-merit}

Even though the penetration depth is a well-trusted measurement for the assessment of sensitivity, the confinement factor is preferred to evaluate single-mode waveguides for bulk sensing, mostly because the lateral evanescent field is neglected in the calculation of $d_\mathrm{p}$. Indeed, a higher penetration depth does not necessarily imply a higher confinement factor, as evident by comparing Table \ref{tab:dp} and Table \ref{tab:width}. From Fig. \ref{fig:FOM} and Table \ref{tab:width}, an inherent compromise between sensitivity to the analyte and waveguide losses was observed: while a larger confinement factor provides a stronger interaction with the analyte, it also makes the waveguide more sensitive to water losses. In fact, when $\Gamma\alpha_\mathrm{H2O}\gg\alpha_\mathrm{int}$, the proposed FOM can be approximated as $\mathrm{FOM}\approx1/\alpha_\mathrm{H2O}$, which yields $\mathrm{FOM}\approx\lambda/(4\pi k_\mathrm{H2O})=4.46\,\mathrm{\upmu m}$ at $\lambda=6$\,µm  (see Eq. \eqref{eq:FOM} and Table \ref{tab:nk}). In this case, the FOM is independent on the specific characteristics of the waveguide and can be interpreted as the distance at which the power of a wave propagating in pure water decays by a factor $1/e$. This is exactly the case for the GOS waveguides and a good approximation for Se6/Se2 waveguides. In the latter, the FOM slightly improves in wider waveguides, as the mode is more confined in the waveguide core and therefore experiences less intrinsic losses due to both absorption from and leakage to the substrate. In the case of GaAs and PCD waveguides, however, intrinsic losses, mainly due to material absorption, become comparable to water-induced losses with increased mode confinement in the waveguide core. Consequently, the FOM exponentially decays with waveguide width. Here, it is worth highlighting that a thicker buffer layer to prevent leakage is only advantageous if the layer is highly transparent at the design wavelength and does not increase losses due to material absorption.  

Selecting a waveguide width among the available range is a degree of freedom in the design. Here, the center of the SM regime was adopted, as this results in a uniform criterion for all the different platforms. Other choices could be, for example, the width at which $\Gamma$ decays to its half-maximum, or that which maximizes the FOM. Practical restrictions, such as the achievable aspect ratio in the fabrication process, must also be considered, especially for PCD waveguides \cite{Malmstrom2016} or platforms with thicker guiding layers, as GaAs. 

\subsection{Limit of detection} \label{subsec:LODdis}

Water absorption significantly limits performance in the 6 µm  wavelength regime. In fact, the optimum lengths in Table \ref{tab:LOD6} are so short that ensuring those are the exact interaction lengths may be challenging during experiments. One practical way to achieve full control of the sensing pathlength is protecting the chip surface with an upper cladding material and lithographically defining a sensing window to uncover the desired fraction of the waveguide \cite{MiBimodal}. As shown in Fig. \ref{fig:LOD} and Table \ref{tab:LOD6}, GaAs waveguides achieve the worst optimum LOD due to their lower sensitivity, paired to comparatively high intrinsic losses. On the other hand, they are very tolerant to longer sensing paths, as they experience significantly less water absorption than the remaining TFWGs. In fact, they become the best choice for sensors longer than 1.8 mm. Se6/Se2 and GOS waveguides achieve a similar FOM (see Table \ref{tab:width}) and, consequently, they exhibit a virtually identical optimum LOD. However, the performance of the Se6/Se2 waveguides quickly degrades with an extended interaction length as they exhibit higher propagation losses, mostly induced by water. PCD waveguides achieve just a slightly worse performance than their GOS counterparts, with a similar tolerance to the sensing pathlength, as they exhibit comparable propagation losses. In the case of GOS, the dominant loss contribution is water absorption, whereas in PCD waveguides, water and intrinsic losses are balanced. If the wavelength is increased up to 10 µm, water absorption decays by nearly 70\%. This is the main reason why longer optimum interaction pathlengths in Table \ref{tab:LOD10} can be derived, especially for GaAs and GOS waveguides, besides improved LODs for all investigated platforms. 

\subsection{Curvature losses} \label{subsec:curvesdis}

Mode mismatch losses are typically the dominant contribution to bending losses and must therefore be carefully addressed to achieve compact, low-loss architectures (see Table \ref{tab:L90}). GaAs waveguides are a paradigmatic example. These waveguides exhibit virtually no radiation losses, but there is a strong mismatch between the straight and the bent mode, yielding the worst radii of Table \ref{tab:L90}. This situation could be reversed by using more sophisticated bending strategies, which are specifically designed to maximize mode overlap in the transition \cite{Hong2022}. As it was to be expected due to their smaller width and superior index contrast, the GOS platform achieves the smallest curves, making it attractive for ultra-compact sensing architectures. PCD waveguides can potentially offer compact curves, especially if mode transition losses are minimized. However, the fabrication of bent diamond SM waveguides remains challenging \cite{Mi2020}. Se6/Se2 offers an intermediate solution, with the advantage that the radius does not scale with the increase in wavelength. Interestingly, for this latter alternative, radiation losses become dominant at $\lambda=10$\,µm. 

\subsection{System-level considerations}

From a systems point-of-view, minimizing the noise in the read-out is key to improving the limit of detection. In fact, order-of-magnitude enhancements in LOD can be accomplished by a judicious SNR optimization \cite{Leuermann2019}. \rv{Considering the noise sources in Eq. (2), and given that low-noise amplifiers are widely available, reducing quantization noise would enable reaching LODs in the shot-noise-limit (see Supp. Mat. S1).} A good practice to reduce $\sigma_\mathrm{DAQ}$ is selecting a DAQ device with a high sampling frequency combined with low-pass filtering or averaging. A high number of bits is also advantageous, likewise adjusting the input signal amplitude such that the entire dynamic range is exploited (see Supp. Mat. S1). Other sources of noise which are not considered here but which may have an impact on the SNR are mechanical \rv{vibrations}, flicker noise or phase noise and \rv{relative intensity noise (RIN)} coming from the light source. It is thus recommended to identify the critical noise contributions on a system-specific basis to ensure optimum working conditions. 

\section{Methods} \label{sec:methods}

Refractive index and absorption data to model the different waveguide materials were extracted from literature \cite{Rakic1996,Skauli2003,Nvemec2014,Shkondin2017,Amotchkina2020,Dore1998,Kischkat2012,Hale1973}. Wavelength dependence of the refractive index (dispersion) and absorption was considered when available. Waveguide modal analysis was performed via 3D FEM simulations using Synopsys’ RSoft FemSIM. A uniform grid of 0.05 µm was employed both for the vertical and horizontal directions. The simulation window included a fraction of the chip substrate and was large enough to assure that the electromagnetic field intensity reaching its borders was negligible. Perfectly Matched Layers (PML) were added to calculate leakage and curvature radiation losses. A PML thickness in excess of 6 µm was selected on a platform-specific basic after carrying out convergence studies. Bent waveguides were simulated with the direct method \cite{Lui1998}. Simulation output was exported to Mathworks’ MATLAB R2020b and processed with in-house scripts. Propagation losses were obtained from the imaginary part of the complex effective index of the mode ($k_\mathrm{eff}$) as $\alpha_\mathrm{wg}=(4\pi/\lambda)k_\mathrm{eff}$. The confinement factor was estimated by evaluating the derivative $\partial k_\mathrm{eff}/\partial k_\mathrm{sample}$, where $k_\mathrm{sample}$ was swept around the value of pure water at the analyzed wavelength \rv{\cite{Gonzalo2019}}. Mode mismatch losses in curves were calculated by the field overlap integral \cite{Subramaniam1997}. 

\rv{\section{General design guidelines}}

\rv{
Our waveguide design approach, described in Section \ref{sec:results} (Results), is not restricted to the selected TFWG platforms, but can be applied generally for the optimization of integrated absorption sensors. For this purpose, we summarize here the key design guidelines.}

\rv{\begin{enumerate}
    \item Identify the critical noise sources and estimate the noise-floor [Eq. \eqref{eq:sigma}] of the target spectroscopic system.
    \item Model the desired TFWG platform at the selected working wavelength in a suitable mode solver, to determine the width range that yields single mode operation and evaluate the confinement factor (Methods), losses [Eq. \eqref{eq:awg}] and FOM [Eq. \eqref{eq:FOM}] in said range.
    \item Select a waveguide width based on a criterion such as sensitivity-loss trade-off, FOM optimization or fabrication constrains.
    \item Evaluate the LOD [Eq. \eqref{eq:LOD}] as a function of interaction pathlength using the noise-floor, confinement factor and losses at the chosen waveguide width. 
    \item Select the waveguide length that minimizes the LOD. If the results are not suitable, consider a different width (point 3) or further optimization of the TFWG platform (point 4).
    \item Evaluate the impact of additional implementation aspects, such as bending losses, in the selected waveguide geometry.
\end{enumerate}}

\section{Conclusions} \label{sec:concl}

In summary, a systematic and rigorous waveguide modelling and design approach has been developed to evaluate different thin-film waveguide platforms toward on-chip mid-infrared spectroscopy and sensing solutions particularly optimized for aqueous analytes. A parametric expression for the LOD of a spectroscopic system (Eq.\,\eqref{eq:LOD}) has been developed, revealing the existence of an optimum interaction length $L_\mathrm{opt}\approx1/\alpha_\mathrm{wg}$. Furthermore, the impact of waveguide geometry on sensitivity and losses was investigated and a global parameter -- $\mathrm{FOM}=\Gamma/\alpha_\mathrm{wg}$ -- accounting for both variables was defined, demonstrating that very different platforms may potentially yield similar results, if they are adequately designed and intrinsic waveguide losses are managed. \rv{Moreover, the FOM is a key factor in the optimum value of the LOD [Eq. \eqref{eq:LODopt}].} The achievable LODs have been compared in terms of optimum performance, governed by the FOM, and tolerance to the interaction pathlength, which is conditioned by the relative importance of solvent absorption within the overall losses. With this study, both a fair comparison between four state-of-the-art TFWG materials, i.e, GaAs, Se6/Se2, GOS and PCD, and a design framework for optimized waveguides were established, which may generically be expanded toward additional platforms. It is anticipated that this rational waveguide design approach will encourage the development of next-generation MIR sensors that readily integrate into lab-on-chip diagnostic platforms. 

\section*{Acknowledgement}
This work was supported by the Ministerio de Universidades, Ciencia e Innovación (FPU19/03330), project TED2021-130400B-I00/AEI/10.13039/501100011033/Unión Europea NextGenerationEU/PRTR and project HORIZON-CL4-2022-DIGITAL-EMERGING-01-03/101093008 (M3NIR).



\end{document}